\documentclass[amsmath, amssymb, twocolumn, floatfix, superscriptaddress, nofootinbib,prl]{revtex4-2}
\usepackage{float}
\usepackage{graphicx}
\usepackage{dcolumn}% Align table columns on decimal point
\usepackage{bm}
\usepackage{color}
\usepackage{xcolor}
\usepackage[colorlinks=true]{hyperref} %Hyperlinks with nice colours
\usepackage[separate-uncertainty=true]{siunitx}
\usepackage{makecell}	
\usepackage{physics}
\usepackage{bbold}
\usepackage{notoccite}
\usepackage[normalem]{ulem}
\usepackage{soul}

\makeatletter
\newcommand{\globalcolor}[1]{%
  \color{#1}\global\let\default@color\current@color
}
\makeatother

\makeatletter
\def\p@subsection{}

\makeatother

\definecolor{lg}{RGB}{200,200,200}
\definecolor{cyberpink}{HTML}{FE53BB}

\newcommand{\figref}[1]{Fig.~\ref{#1}}

\definecolor{teocol}{rgb}{0.9, 0.61, 0.04}

\newcommand{\doubleket}[1]{|#1\rangle\hspace{-2pt}\rangle}
\newcommand{\doublebra}[1]{\langle\hspace{-2pt}\langle#1|}

\newcommand{\PRLsep}{\noindent\makebox[\linewidth]{\resizebox{0.625\linewidth}{1pt}{$\bullet$}}\bigskip}

\begin{document}

\title{Demonstration of a quantum SWITCH in a Sagnac configuration}

\author{Teodor Strömberg}\email[Corresponding author: ]{teodor.stroemberg@univie.ac.at}
\author{Peter Schiansky}\affiliation{University of Vienna, Faculty of Physics \& Vienna Doctoral School in Physics,  Boltzmanngasse 5, A-1090 Vienna, Austria}
\affiliation{University of Vienna, Faculty of Physics \& Research Network Quantum Aspects of Space Time (TURIS), Boltzmanngasse 5, 1090 Vienna, Austria}
\author{Robert W. Peterson}\affiliation{University of Vienna, Faculty of Physics \& Research Network Quantum Aspects of Space Time (TURIS), Boltzmanngasse 5, 1090 Vienna, Austria}
\author{Marco Túlio Quintino}
\affiliation{Sorbonne Universit\' {e}, CNRS, LIP6, F-75005 Paris, France}
\affiliation{University of Vienna, Faculty of Physics, Boltzmanngasse 5, 1090 Vienna, Austria}
\affiliation{Institute for Quantum Optics and Quantum Information, Boltzmanngasse 3, 1090 Vienna, Austria}
\author{Philip Walther}\email[Corresponding author: ]{philip.walther@univie.ac.at}\affiliation{University of Vienna, Faculty of Physics \& Research Network Quantum Aspects of Space Time (TURIS), Boltzmanngasse 5, 1090 Vienna, Austria}

\date{\today}

\begin{abstract}
The quantum SWITCH is an example of a process with an indefinite causal structure, and has attracted attention for its ability to outperform causally ordered computations within the quantum circuit model. To date, realisations of the quantum SWITCH have made a trade-off between relying on optical interferometers susceptible to minute path length fluctuations and limitations on the range and fidelity of the implementable channels, thereby complicating their design, limiting their performance and posing an obstacle to extending the quantum SWITCH to multiple parties. In this Letter we overcome these limitations by demonstrating an intrinsically stable quantum SWITCH utilizing a common-path geometry facilitated by a novel reciprocal and universal $\mathrm{SU}(2)$ polarization gadget. We certify our design by successfully performing a channel discrimination task with near unity success probability.
\end{abstract}

\maketitle

\textit{Introduction}---
Quantum information processing tasks are most commonly described within the framework of the quantum circuit model. In this framework an initial state gradually evolves by passing through a fixed squence of gates. This, however, is not the most general model of computation that quantum mechanics admits, and in~\cite{chiribella2013quantum} a processes that effects a superposition of quantum circuits was proposed. This process, known as the quantum SWITCH, has attracted significant theoretical~\cite{colnaghi2012quantum,chiribella2012perfect,feix2015quantum,zhao2020quantum} and experimental~\cite{procopio2015experimental,rubino2017experimental,goswami2018indefinite,rubino2019experimental,goswami2020increasing,rubino2021experimental,guo2020experimental,cao2022experimental} interest. Together with the so-called Oreshkov-Costa-Brukner process~\cite{oreshkov2012quantum} it was the first example of a quantum process without a definite causal structure, and motivated the study of more general causal structures within quantum mechanics that could help bridge the gap between general relativity and quantum mechanics. The quantum SWITCH is also of practical interest, since it has been shown to allow for a computational advantage over standard quantum circuits~\cite{Araujo2014,renner2022computational}, an advantage which has been demonstrated experimentally~\cite{wei2019experimental}.

In its simplest form, the quantum SWITCH is a map that acts on two gates, $U$ and $V$, and transforms them into a controlled superposition of the gates being applied in two different orders:
\begin{equation}
    (U,V) \mapsto
    UV\otimes \ketbra{0}{0}_C + VU\otimes \ketbra{1}{1}_C.
    \label{eq:2switch}
\end{equation}
\begin{figure}[t]
    \includegraphics[width=1\linewidth]{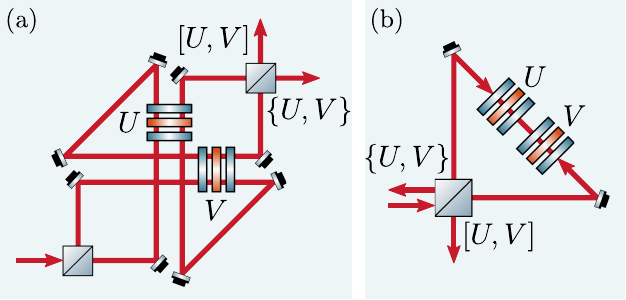}
    \centering
    \caption{\textbf{Path-polarization quantum SWITCH.} \textbf{(a)} The most common implementation of the photonic quantum SWITCH utilizes the path degree of freedom of a single photon inside a Mach-Zehnder interferometer to coherently control the order in which two polarization operations $U$ and $V$ are applied. In this geometry, photons in different arms of the interferometer propagate through different parts of the polarization optics. \textbf{(b)} An implementation based on a Sagnac interferometer is fundamentally simpler and more robust, but necessitates two different propagation directions through the polarization gadgets effecting the transformations $U$ and $V$. In general the operations in the two different propagation directions are not the same, limiting the use of this geometry to special cases.}
    \label{fig:mzi}
\end{figure}
\begin{figure*}[t]
    \includegraphics[width=1.00\linewidth]{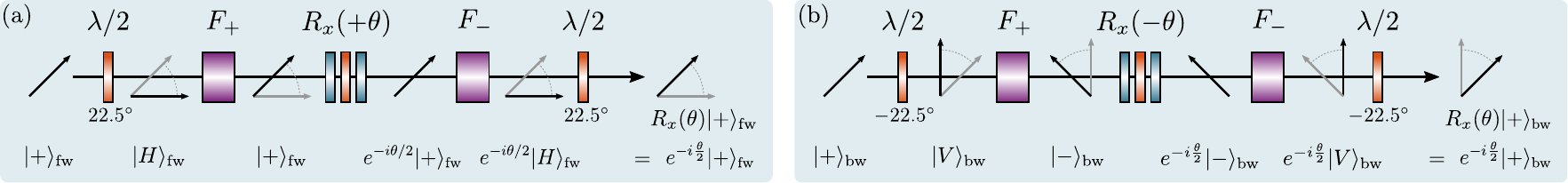}
    \centering
    \caption{\textbf{Reciprocal polarization gadget.} 
    The evolution of a diagonally polarized state in the two different propagation directions through a reciprocal $R_x$ gadget. The arrows indicating the polarization states are drawn in the co-moving frame. Note that only differential phases between the two polarization components are indicated. \textbf{(a)} The Faraday rotators $F_{\pm}$ rotate linear polarization by \SI{45}{\degree}, and in the forwards direction these rotations cancel the half-wave plate (HWP) rotations, indicated by $\lambda/2$. The state before the central waveplate gadget is therefore $\ket{+}$, and the correct phase gets applied. \textbf{(b)} In the backwards propagation direction the rotations of the HWP and Faraday rotators add, rotating the linear polarization by \SI{90}{\degree}. The $R_x(-\theta)$ rotation applies the desired phase to what is now the $\ket{-}$ state, which subsequently gets rotated back to $\ket{+}$, yielding the correct transformation.
    }
    \label{fig:gadget}
\end{figure*}
To date, all experimental realisations of the quantum SWITCH have been done using single photons as the physical system encoding the input and output state of the process. These implementations typically rely on folded Mach-Zehnder interferometers (MZIs) and polarization optics to couple different internal degrees of freedom of the single photons. A technical challenge associated with such implementations is that the phase of the interferometer needs to be kept constant even as different choices of $U$ and $V$ in \eqref{eq:2switch} change the interference condition.  In practice, most experimental quantum SWITCHes have relied on passive phase stability during operation, limiting not only their fidelity, but also their duty cycle due to the need to periodically reset the phase. Furthermore, the geometry of the MZI means that single photons in the two different arms of the interferometer interact with different parts of the polarization optics  (see \figref{fig:mzi}a). The reliance on optical geometries that suffer from phase instability is necessitated by the non-reciprocity of the optical components that effect the unitary transformations $U$ and $V$ on the photon polarization. The    works~\cite{goswami2018indefinite,goswami2020increasing} remedied this by using the polarization as a control degree of freedom (DOF), instead of a target, thereby enabling a common-path geometry, however this came at the expense of more cumbersome and low fidelity target qubit operations. 

In~\cite{wei2019experimental} a common-path geometry was realised in a different way, by encoding the target system in the temporal degree of freedom. This implementation, however, was limited to generalized versions of the Pauli $X$ and $Z$ operators, and could therefore not prepare superposition states. These operations furthermore required both ultra-fast phase modulators and the manual replacement of optical components, increasing the experimental requirements while reducing the programmability of the setup.

In this Letter we overcome these limitations by designing a fully reciprocal polarization gadget capable of realising any $U\in \mathrm{SU}(2)$, thereby enabling the use of a passively stable Sagnac geometry with perfect spatial mode overlap, while still using the polarization DOF for the target qubit, without imposing any restrictions on the unitaries applied on this system inside the quantum SWITCH.

\textit{Photonic quantum SWITCH}---
In the commonly used path-polarization quantum SWITCH, shown in \figref{fig:mzi}a, the path DOF is used to coherently superpose two different orders through the waveplate gadgets that act on the target (polarization) qubit. These waveplate gadgets, first introdued by Simon and Mukunda, consist of two quarter-wave plates and one half-wave plate, and are universal for $\mathrm{SU}(2)$~\cite{simon1990minimal}. If one tries to use a common path geometry, as depicted in \figref{fig:mzi}b, the photon now travels through the polarization gadgets in two different directions. The action of a Simon-Mukunda gadget in the backwards direction is: $U^{\text{bw}} = P U^T P^{\dagger} \neq U$, where $U$ is the unitary operation in the forwards direction, and $P$ is a unitary operator describing the basis change to the backwards propagation direction. By picking a convention for the polarization states in which the diagonal polarizations are associated with the eigenstates of the Pauli $Y$ matrix one finds that $P = \mathbb{1}$~\cite{stromberg2022experimental}, however the residual transpose, a consequence of the fact that the order of the waveplates is transposed in the backwards direction, cannot be undone this way. The Simon-Mukunda gadget is therefore only reciprocal for a two parameter subset of $\mathrm{SU}(2)$, and common-path quantum SWITCHes such as~\cite{schiansky2023demonstration}, were thus far not able to implement arbitrary polarization unitaries.
\begin{figure*}[t]
    \includegraphics[width=1\linewidth]{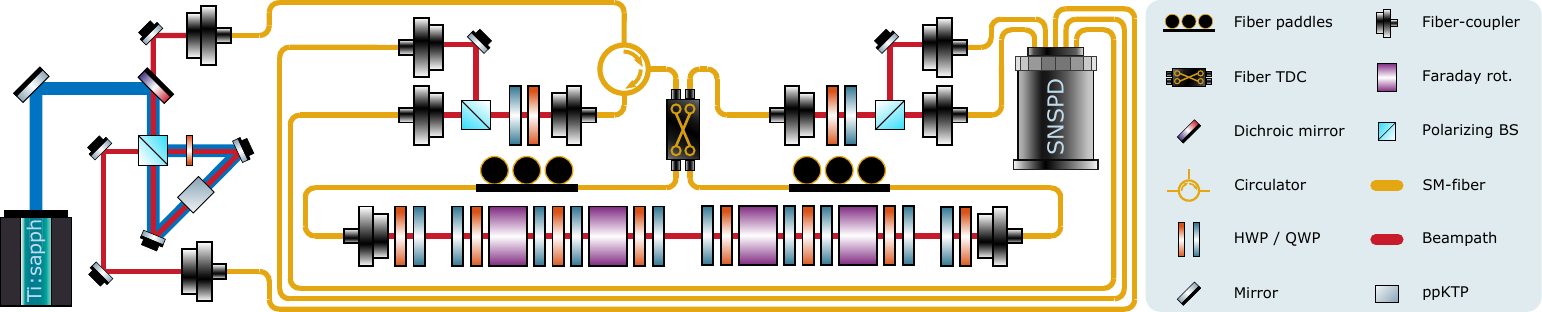}
    \centering
    \caption{\textbf{Experimental setup.} Single photons are generated by a type-II spotaneous parametric down-conversion source using a ppKTP crystal. Detection of the signal photon using superconducting nanowire single-photon detectors (SNSPDs) heralds the presence of the idler photon. A tunable directional coupler (TDC) configured for a balanced splitting ratio sends the idler photon through a free-space path in a superposition of two propagation directions. This path contains two reciprocal polarization gadgets consisting of Faraday rotators, quarter-wave plates (QWPs) and half-wave plates (HWPs).
    %Out of these seven waveplates, two quarter-wave plates (QWPs) and three half-wave plates (HWPs) are motorized.
    These gadgets implement the operators $U$ and $V$. The photon finally exits in one of the two TDC ports depending on whether $U$ and $V$ commute or anti-commute, and is then detected using a polarization resolving measurement. A fiber circulator is used to pick off photons exiting the Sagnac in the input port.}
    \label{fig:setup}
\end{figure*}

In this work, we adopt the convention $(S_1,S_2,S_3) \leftrightarrow (X,Y,Z)$ for the Stokes parameters and Pauli matrices. Written in this convention, the Simon-Mukunda gadget transforms as: $U \mapsto U^{\text{{bw}}} = ZU^T Z$ under counter propagation. Given a unitary parameterised as: $U=\mathrm{exp}[-i\frac{\theta}{2}\vec{\sigma}\cdot\vec{n}]$, where $\vec{\sigma}$ is the Pauli vector and $\vec{n}$ the rotation axis of the unitary operation on the Bloch sphere, this transformation corresponds to $\begin{bmatrix}
    \theta & n_x & n_y & n_z
    \end{bmatrix}
    \mapsto
    \begin{bmatrix}
    \theta & -n_x & n_y & n_z
    \end{bmatrix}$.
Since this transformation applies to any unitary operation implemented by a sequence of linear retarders under counter propagation, we also consider circular retarders, more specifically Faraday rotators. It is a well known fact that Faraday rotators are non-reciprocal, due to the magneto-optic effect breaking Lorentz-reciprocity~\cite{asadchy2020tutorial}. This property has enabled a multitude of widely adopted optical devices such as Faraday mirrors~\cite{martinelli1989universal}, optical circulators, and optical isolators~\cite{saleh2019fundamentals}. Quantitatively, the non-reciprocity manifests itself as the following transformation under counter propagation $[\theta \;\; n_y] \mapsto [-\theta \;\; n_y]$.
    
Faraday rotators are usually sold with a fixed circular retardance of $\theta = \frac{\pi}{2}$, corresponding to a rotation of linear polarization by \SI{45}{\degree}. We will therefore restrict our discussion to only these devices, and show how they can be used to construct a fully reciprocal polarization device.

\textit{A reciprocal polarization gadget}---
Since only the $X$-component of the Simon-Mukunda gadget exhibits non-reciprocity, we begin by constructing a gadget capable of realising a reciprocal $X$-rotation:
\begin{equation}
    G_x(\theta) = 
    H\Big(\frac{\pi}{8}\Big)
    F_-
    Q\Big(\frac{\pi}{2}\Big)
    H\bigg(\frac{\theta+2\pi}{4}\bigg)
    Q\Big(\frac{\pi}{2}\Big)
    F_+
    H\Big(\frac{\pi}{8}\Big).
\end{equation}
Here $H$ and $Q$ refer to half- and quarter-wave plates at a given angle from the vertical axis, and $F_{\pm}$ are Faraday rotators with circular retardance of $\pm \frac{\pi}{2}$. The three middle waveplates constitute a Simon-Mukunda gadget implementing a non-reciprocal $X$-rotation:
\begin{equation}
R_x(\theta+2\pi) = {-R_x}(\theta) \mapsto {-R_x}(-\theta).
\end{equation} The action of the $G_x$ gadget in the two different propagation directions can therefore be expressed as:
\begin{align}
    G_x^{\text{fw}}(\theta) &= 
    -H\Big(\frac{\pi}{8}\Big)
    F_-
    R_x(\theta)
    F_+
    H\Big(\frac{\pi}{8}\Big) \\
    G_x^{\text{bw}}(\theta) &=
    -H\Big(\!\!-\frac{\pi}{8}\Big)
    F_-
    R_x(-\theta)
    F_+
    H\Big(\!\!-\frac{\pi}{8}\Big),
\end{align}
since for a single linear retarder at an angle $\varphi$ to the vertical axis the effect of reversing the propagation direction is: $\varphi \mapsto -\varphi$. The superscripts `$\text{fw}$' and `$\text{bw}$' refer to the forwards and backwards propagation directions respectively. To see that the full gadget is reciprocal, note that
\begin{align}
    H\Big(\frac{\pi}{8}\Big)F_- &= F_+ H\Big(\frac{\pi}{8}\Big) = -iX \\
    H\Big(\!\!-\frac{\pi}{8}\Big)F_- &= F_+H\Big(\!\!-\frac{\pi}{8}\Big) = -iZ,
\end{align}
as shown in the Supplementary Material. The action of the gadget in the two propagation directions can therefore be simplified to:
\begin{align}
    G_x^{\text{fw}}(\theta) &= X R_x(\theta) X = R_x(\theta) \\
    G_x^{\text{bw}}(\theta) &= Z R_x(-\theta) Z = R_x(\theta).
\end{align}
A graphical examination of the reciprocity of the gadget is shown in \figref{fig:gadget}. Using this gadget as a building block, it becomes possible to construct a fully reciprocal gadget capable of implementing arbitrary unitaries:
\begin{equation}
\label{eq:igadget}
    G_R = Q(\theta)H(\phi)G_x(\gamma)H(-\phi)Q(-\theta).
\end{equation}
This gadget is reciprocal due to its palindromic order, and a proof of its universality, as well as a method to find the angles $\theta$, $\phi$ and $\gamma$ for a given $U$, is provided in the Supplementary Material. An implementation of this algorithm is available in an open repository~\cite{MTQgit}.

\textit{Advantage in a channel discrimination task}---
To certify our that our experimental platform is capable of realising an indefinite causal order, we now present a channel discrimination task for which the quantum SWITCH strictly outperforms any causally ordered strategy. This channel discrimination problem was originally presented as a causal witness in Ref.~\cite{Araujo2014} and was inspired by the task introduced in Ref.~\cite{chiribella2012perfect}. Let $U_i,V_j$ be two qubit unitary operators belonging to the set
\begin{equation}
    \mathcal{G} := \left\{ \mathbb{1}, X, Y, Z, \frac{X\pm Y}{\sqrt{2}}, \frac{X\pm Z}{\sqrt{2}}, \frac{Y\pm Z}{\sqrt{2}} \right\}
    \label{eq:gset}.
\end{equation}
\begin{figure*}[t]
    \includegraphics[width=1.00\linewidth]{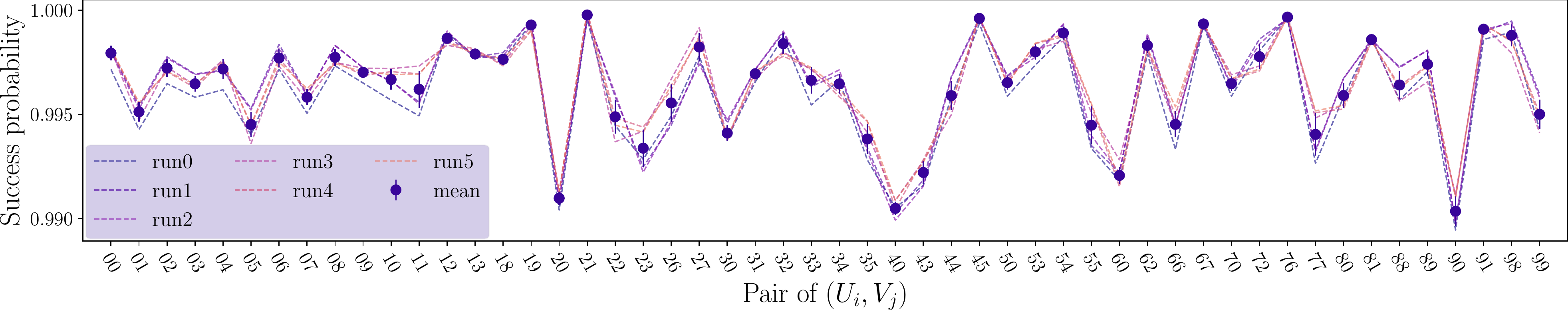}
    \centering
    \caption{%\textbf{Success probabilities.}
    Success probabilities $p_{\text{s}}(i,j)$, which correspond to the probability for the photon to exit in the correct port of the interferometer given a pair of commuting or anti-commuting unitaries $(U_i,V_j)$. The six different runs of the experiment, plotted separately, exhibit high repeatability. Dashed lines show these six different experimental runs, while the solid dots indicate the mean and standard deviation of the runs. The average success probability is $\langle p_{\text{s}}\rangle = 0.9964$, while the highest and lowest success probabilities are $\mathrm{max}\bigl(p_{\text{s}}(i,j)\bigr) = 0.99997$ and $\mathrm{min}\bigl(p_{\text{s}}(i,j)\bigr) = 0.9895$ respectively.}
    \label{fig:winning_probs}
\end{figure*}

Using this set, we define two sets of pairs of operators $(U_i,V_j)$ that either commute or anti-commute:
\begin{align}
    \label{eq:setgpm}
     &\mathcal{G}_{\pm}:=\left\{\left(U_i,V_j\right) \Big| U_i,V_j \in \mathcal{G}, \; U_iV_j =\pm V_jU_i  \right\}
\end{align}
Let $(U_i,V_j)$ be a pair of channels belonging to either $\mathcal{G}_+$ or $\mathcal{G}_-$, and consider the task of deciding to which set they belong, given only a single use of the channels. It is well known that this task can be performed deterministically when given access to the quantum SWTICH. This can be seen by setting the state of the control qubit in \eqref{eq:2switch} to $\ket{+}_C$ and considering the action on any target state $\ket{\Psi}_T$:
\small
\begin{equation}
\begin{aligned}
&\left( UV\otimes\ketbra{0}{0} +VU\otimes\ketbra{1}{1}\right) \ket{\Psi}_T\otimes\ket{+}_C  =\\ 
&    \frac{1}{2}(UV+VU)\ket{\Psi}_T\otimes\ket{+}_C + 
    \frac{1}{2}(UV-VU)\ket{\Psi}_T\otimes\ket{-}_C.
\end{aligned}
\end{equation}
\normalsize
A measurement of the control qubit then reveals to which set  $(U_i,V_j)$ belongs.

For this discrimination task, the probability of successfully guessing the set can be expressed as
\begin{equation}p_{\text{s}}(i,j) := 
        p\bigl(\pm|(U_i,V_j)\bigr),\quad \text{ if } (U_i,V_j)\in \mathcal{G}_{\pm}
\end{equation}
By making use of the semidefinite programming methods presented in Ref.~\cite{bavaresco21}, we find that any causally ordered strategy necessarily obeys: $\min\bigl(p_{\text{s}}(i,j)\bigr)\leq 0.841.$
Moreover, if the pairs of channels $(U_i,U_j)$ are uniformly picked from $\mathcal{G}_+$ and $\mathcal{G}_-$, the average  probability of correctly guessing the set with a causally ordered strategy is bounded by $ \frac{1}{N}\sum_{i,j}p_{\text{s}}(i,j)\leq 0.904$ where the indices $i,j$ run over all the pairs of gates that commute or anti-commute. As discussed in Ref.~\cite{Araujo2014}, this average success probability approach can be phrased in terms of a causal non-separability witness, and in the Supplementary Material we explicitly present such a witness. 

\textit{Experiment}---
Before experimentally performing the channel discrimination task, we first show that the gadget in \eqref{eq:igadget} is indeed reciprocal and universal. %capable of implementing all the required unitaries. 
To this end we perform quantum process tomography on both gadgets used in the experiment for 100 random unitaries. The resulting gate fidelities, defined as the average state fidelity under the reconstructed unitaries, are presented in the Supplementary Material. Achieving an average gate fidelity of \num{0.9970\pm0.0018}, and an average fidelity between the two propagation directions of \num{0.9972\pm0.0018}, we conclude that the gadget is reciprocal and universal.

We now turn to the experimental realisation of the quantum SWITCH, pictured in \figref{fig:setup}. For the certification of the indefinite causal structure of the implemented process, we employ single photons generated using type-II spontaneous parametric down-conversion~\cite{greganti2018tuning}. The signal photon is used as a herald for the idler photon, which is made to propagate through the quantum SWITCH. Initially, a beam-splitter in the form of a tunable directional coupler (TDC) applies a Hadamard operation on the path (control) DOF, thereby preparing the state: $\ket{\Psi}_T\otimes (\ket{0}_C + \ket{1}_C)/\sqrt{2}$,
where the subscripts $C$ and $T$ refer to the control and target degrees of freedom respectively. The photon then passes through the polarization gadgets in a superposition of the two propagation directions, correlating the applied gate order with the control DOF: $    \frac{1}{\sqrt{2}}UV\ket{\Psi}_T\otimes \ket{0}_C +
    \frac{1}{\sqrt{2}}VU\ket{\Psi}_T\otimes \ket{1}_C.$
Finally, the photon propagates back to the TDC which once again applies a Hadamard gate on the control qubit:
\begin{equation}
    \frac{1}{2}\{U,V\}\ket{\Psi}_T\otimes\ket{0}_C +
    \frac{1}{2}[U,V]\ket{\Psi}_T\otimes\ket{1}_C.
\end{equation}
Measuring the photon's location then reveals, with unity probability, whether the gates $(U,V)$ commute or anti-commute. 

Our implementation makes use of a combination of free-space and fiber optics, which is facilitated by the intrinsic phase stability of the common-path geometry. The use of a fiber TDC allows for precise control over the splitting ratio as well as providing perfect spatial mode overlap. These two factors combine to yield a high interferometric visibility in excess of $0.9995$. The two inner ports of the TDC are connected to fiber collimators that launch the photons into free space where they propagate through the two polarization gadgets in opposite directions. 

A fiber-circulator is placed at the input port of the TDC to separate the backwards propagating photons from the input light. Finally, two measurement stations are used to measure the polarization of the photons in either output arm of the Sagnac. The polarization resolving measurements allow for the polarization dependent detection efficiencies in the superconducting nanowire single-photon detectors used in the experiment to be corrected for. The fiber circulator induces a small amount of differential loss in the two interferometer outputs which is also corrected for (see Supplementary Material).

\textit{Results}---
Each of the 52 pairs of (anti-)commuting unitary operators in the sets \eqref{eq:setgpm} were implemented six independent times, and for each pair of operations single-photon events were recorded for 60 seconds, giving a total measurement time of approximately \SI{5}{\hour}, with the only downtime being the time spent rotating the waveplates. 
The success probabilities were then calculated separately for each run. The results of this are shown in \figref{fig:winning_probs}. We find a minimum success probability of $\min \bigl(p_{\text{s}}(i,j)\bigr) = 0.9895$, and an average success probability of $\langle p_{\text{s}}\rangle=0.99639\pm0.00007$, far exceeding the causally separable bounds of $0.841$ and $0.904$, respectively. Our observed average success probability can be directly compared with a non-common-path implementation of an analogous channel discrimination task presented in~\cite{procopio2015experimental}, where a success probability of $0.973$ was achieved. The observed success probabilities $p_{\text{s}}(i,j)$ for the individual pairs of gates $(U_i,V_j)$ additionally display a remarkably low variance, with a recorded standard deviation of $\sigma^2 = 0.0024$, demonstrating the robustness of our design. The uncertainty in the experimentally evaluated success probability is the error-propagated observed standard deviation for the constituent success probabilities $p_{\text{s}}(i,j)$ in the six runs.

\textit{Discussion}---
We have demonstrated for the first time a path-polarization quantum SWITCH that utilizes a passively stable common-path geometry. Our novel design greatly simplifies the construction and operation of the device, while simultaneously increasing its fidelity, robustness and duty cycle compared to previous demonstrations. The implementation is facilitated by a new polarization gadget that combines different forms of non-reciprocity to unlock fully reciprocal and universal polarization transformations. The methods used to engineer the reciprocity of the polarization gadgets can be applied more broadly to map balanced interferometers onto common-path geometries. This will enable simple and robust bulk-optics realisations of important primitives such as reconfigurable beam-splitters and variable partially-polarizing beam-splitters~\cite{florez2018variable}. We anticipate that this will lead to straightforward realisations of generalized measurements directly on polarization qubits~\cite{kurzynski2013quantum}, as well as the demonstration of multi-party quantum SWITCHes~\cite{renner2022computational}.
\\
%The simplicity of our approach will aid in the construction of higher order quantum SWITCHes, which have so far not been realised due to the complexity of the associated phase stabilisation. We also anticipate that the polarization gadget presented here will have applications outside the field of indefinite causality, for example in the realisation of passively stable and tunable beam-splitters and partially polarizing beam-splitters based on common-path interferometers~\cite{florez2018variable}, as well as implementations of generalized measurements on polarization qubits~\cite{kurzynski2013quantum}.
%\\
\begin{acknowledgments}
    T.S. thanks Francesco Massa for helpful discussions. R.W.P. acknowledges support from the ESQ Discovery program (Erwin Schr\"{o}dinger Center for Quantum Science and Technology), hosted by the Austrian Academy of Sciences (\"{O}AW).  P.W. acknowledges support from the research platform TURIS, the European Commission through EPIQUS (no. 899368) and AppQInfo (no. 956071). from the Austrian Science Fund (FWF) through BeyondC (F7113) and Reseach Group 5 (FG5), from the AFOSR via PhoQuGraph (FA8655-20-1-7030) and QTRUST (FA9550-21- 1-0355), from the John Templeton Foundation via the Quantum Information Structure of Spacetime (QISS) project (ID 61466), and from the Austrian Federal Ministry for Digital and Economic Affairs, the National Foundation for Research, Technology and Development and the Christian Doppler Research Association.
\end{acknowledgments}

All computational code used in this work is openly available at~\cite{MTQgit}. 

%\bibliography{bibliography}

\onecolumngrid
\PRLsep
\appendix
\section{Supplementary material}
\subsection{Experimental details}
The single photons in the experiment were generated using spontaneous parametric down-conversion in a periodically poled $\mathrm{KTiOPO_4}$-crystal phase matched for a type-II collinear process. This crystal was pumped by a pulsed Ti:Sapphire laser (Coherent Mira 900HP), with a pulse repetition rate of \SI{76}{\mega\hertz} and tuned to a wavelength of $\lambda_p = \SI{773}{\nano\meter}$, thereby generating degenerate photon pairs at $\lambda_s = \lambda_i = \SI{1546}{\nano\meter}$. The two photons were separated using a polarizing beam-splitter, and the signal photon was sent directly to a superconducting nanowire single-photon detector (SNSPD) from PhotonSpot, housed in a \SI{1}{\kelvin} cryostat. These single-photon detectors had a detection efficiency of around \SI{95}{\percent}, and were separated from the experimental setup by \SI{100}{\meter} of optical fiber. The idler photon was sent to the experimental setup through approximately \SI{10}{\meter} of single-mode fiber, and was then injected into the tunable directional coupler (TDC) using a fiber-optic circulator. This circulator contributed approximately \SI{1}{\decibel} of optical loss per pass (\SI{2}{\decibel} total).

Two \SI{5}{\meter} fibers spooled inside fiber polarization controllers were used to connect the output of the TDC to the fiber couplers in the centre of the Sagnac. The free-space optical path loop was approximately \SI{80}{\centi\meter} long, in order to have sufficient room for all the polarization optics and leaving enough space to fit polarizers between the elements for characterisation measurements. Due to the small \SI{5}{\milli\meter} aperture of the Faraday rotators in the polarization gadgets, fiber collimators producing a small beam-diamter (Thorlabs PAF2A-7C) were used to ensure the spatial profile of the photons was not clipped by the polarization elements. The single-mode coupling efficiency in the free-space part was in excess of \SI{85}{\percent}. In order to reduce backreflections in the interferometer, the fibers connected to the fiber collimators used anti-reflection coated APC connectors. The output ports of the TDC in the backwards direction were connected to two polarization measurement stations, and the photons in the TDC output port overlapping with the input port were separated by the fiber circulator. The two different polarization components of the light were separated using polarizing beam-splitters and coupled into different single-mode fibers, connected to a total of four SNSPDs. This was done in order to account for the polarization dependent detection efficiencies in these detectors.

Before performing the measurements described in the main text, polarization compensation was first performed on the fibers inside the Sagnac. This was done by injecting  $H$ ($+$)-polarized CW light in the input port of the interferometer, and using the fiber paddles and quarter- / half-wave plates next to the fiber collimators to minimize the transmission through a $V$ ($-$)-polarizer. Polarization contrasts in excess of \SI{40}{\decibel} were achieved for both input polarization states. As a final step, the splitting ratio of the TDC was finetuned. This was done by configuring both polarization gadgets to implement the identity operation, such that destructive interference is observed in one output port of the interferometer. The splitting ratio of the TDC was optimized by minimizing the optical power in this dark port.

\subsection{Definitions and conventions}
In this work we use the following convention for our polarization states:
\begin{equation}
    \begin{alignedat}{3}
        &\ket{H} = \begin{bmatrix}
            1\\0
        \end{bmatrix},\qquad
        &&\ket{V} = \begin{bmatrix}
            0\\1
        \end{bmatrix},
        \\
        &\ket{+} = \frac{1}{\sqrt{2}}\begin{bmatrix}
            1\\1
        \end{bmatrix},\qquad
        &&\ket{-} = \frac{1}{\sqrt{2}}\begin{bmatrix}
            1\\-1
        \end{bmatrix},
        \\
        &\ket{L} = \frac{1}{\sqrt{2}}\begin{bmatrix}
            1\\i
        \end{bmatrix},\qquad
        &&\ket{R} = \frac{1}{\sqrt{2}}\begin{bmatrix}
            1\\-i
        \end{bmatrix}.
    \end{alignedat}
\end{equation}
Under this convention, quarter-wave and half-wave plates are defined as:
\begin{align}
    Q(\theta) &= R_y(2\theta)R_z(\pi/2)R_y(-2\theta)\\
    H(\theta) &= R_y(2\theta)R_z(\pi)R_y(-2\theta),
\end{align}
where
\begin{equation}
    R_k(\theta) = \mathrm{exp}\biggl[
    -i\frac{\theta}{2}\sigma_k
    \biggr] = \cos\frac{\theta}{2} I -i\sin\frac{\theta}{2} \sigma_k.
\end{equation}
Similarly, the fixed Faraday rotators are expressed as:
\begin{equation}
    F_{\pm} = R_y(\pm \pi/2).
\end{equation}
\subsection{Gadget derivation}
Having established these definitions, we explicitly show the simplification used in the derivation of the reciprocal gadget in the main text:
\begin{equation}
\begin{aligned}
    H\Bigl(\frac{\pi}{8}\Bigr)
    F_- &=
    R_y\Bigl(\frac{\pi}{4}\Bigr) 
    R_z(\pi)
    R_y\Bigl(-\frac{\pi}{4}\Bigr)
    R_y\Bigl(-\frac{\pi}{2}\Bigr)\\
    &=
    R_z(\pi)
    R_y\Bigl(-\frac{\pi}{4}\Bigr) 
    R_y\Bigl(-\frac{\pi}{4}\Bigr)
    R_y\Bigl(-\frac{\pi}{2}\Bigr)\\
    &=
    R_z(\pi) R_y(-\pi)\\
    &= (-iZ)(iY)\\
    &= -iX,
    \end{aligned}
\end{equation}
where the first step used:
\begin{equation}
    R_y(\theta) R_z(\pi)= R_z(\pi)R_y(-\theta).
\end{equation}
The other three simplifications follow using the same steps.

\subsection{Universality of the reciprocal gadget}
In this section we will give a proof that the reciprocal gadget presented in the main text is capable of implementing any $U\in \mathrm{SU}(2)$. We first recall the construction of this gadget:
\begin{equation}
\begin{aligned}
\label{eq:gadetappendix}
    G_R &= Q(\theta)H(\phi)G_x(\psi)H(-\phi)Q(-\theta) \\
    &= Q(\theta)H(\phi)R_x(\psi)H(-\phi)Q(-\theta).
    \end{aligned}
\end{equation}
In~\cite{simon1990minimal} it was shown that a combination of one half-wave and quarter-wave plate implements a two-parameter subset of $\mathrm{SU}(2)$ parameterized as:
\begin{equation}
    Q(\theta)H(\phi) = R_y(\alpha) R_z(\pi/2) R_y(\beta).
\end{equation}
This subset can be equivalently expressed as:
\begin{equation}
    R_y(\gamma) R_z(\delta) R_x(\pi/2)
    \iff R_y(\alpha) R_z(\pi/2) R_y(\beta).
\end{equation}
Given a two-waveplate gadget with the above parameterisation, the description in the backwards propagation direction is:
\begin{equation}
    H(-\phi)Q(-\theta) = R_x(-\pi/2) R_z(\delta) R_y(\gamma).
\end{equation}
Substituting in these parameterisations in \eqref{eq:gadetappendix} we find:
\begin{equation}
    \begin{aligned}
    G_R &= R_y(\gamma) R_z(\delta) R_x(\pi/2)
    R_x(\psi)
    R_x(-\pi/2) R_z(\delta) R_y(\gamma) \\
    &=R_y(\gamma) R_z(\delta)
    R_x(\psi)
    R_z(\delta) R_y(\gamma).
    \end{aligned}
\end{equation}
We now multiply this expression from the left by $R_x(\pi)$ and use the trivial relation $R_x(\pi) R_x(-\pi) = \mathbb{1}$:
\begin{equation}
\label{eq:xtimesgadget}
    \begin{aligned}
    R_x(\pi) G_R
    &= R_x(\pi) R_y(\gamma)R_x(-\pi)
    R_x(\pi)R_z(\delta)
    R_x(\psi)
    R_z(\delta) R_y(\gamma) \\
    &= R_x(\pi) R_y(\gamma)R_x(-\pi)
    R_x(\pi)R_z(\delta)R_x(-\pi)
    R_x(\psi+\pi)
    R_z(\delta) R_y(\gamma)\\
    &= R_y(-\gamma)R_z(-\delta)R_x(\psi+\pi)
    R_z(\delta) R_y(\gamma)\\
    &= R_y(-\gamma)R_z(-\delta)R_x(\psi')
    R_z(\delta) R_y(\gamma),
    \end{aligned}
\end{equation}
where $\psi' = \psi+\pi$ and we made use of the identities:
\begin{align}
    R_x(\pi) R_y(\gamma)R_x(-\pi) &= R_y(-\gamma) \\
    R_x(\pi) R_z(\delta)R_x(-\pi) &= R_z(-\delta),
\end{align}
in the last step. To show that \eqref{eq:xtimesgadget} is universal it suffices to show that it can apply a phase $\lambda/2$ to an arbitrary state $\ket{u}$:
\begin{equation}
    R_x(\pi) G_R \ket{u} = U\ket{u} = e^{i\lambda/2}\ket{u}.
\end{equation}
To this end, we choose $\psi'$, $\delta$ and $\gamma$ such that $R_x(\psi')R_z(\delta) R_y(\gamma)$ maps $\ket{u}$ to $\ket{-}$ times some phase $\phi$:
\begin{equation}
    R_x(\psi')R_z(\delta) R_y(\gamma) \ket{u} = e^{i\phi}\ket{-}=e^{i(\psi' + \mu)/2}\ket{-},
\end{equation}
where $\mu = \phi - \psi'$. This is always possible since $R_x(\psi')R_z(\delta) R_y(\gamma)$ is a Tait-Bryan rotation. It's evident that the mapping $\ket{v}\rightarrow e^{i\mu/2}\ket{-}$ has to be done by $R_z(\delta) R_y(\gamma)$, since $\ket{-}$ is an eigenstate of $R_x(\psi')$:
\begin{equation}
\begin{aligned}
    R_z(\delta) R_y(\gamma)\ket{u} &=  e^{i\mu/2}\ket{-}\\
    R_x(\psi')\ket{-} &= e^{i\psi'/2} \ket{-}.
\end{aligned}
\end{equation}
We therefore have:
\begin{equation}
\begin{aligned}
    R_x(\pi) G_R \ket{u} &= R_y(-\gamma)R_z(-\delta)R_x(\psi')
    R_z(\delta) R_y(\gamma)\ket{u} \\
    &= R_y(-\gamma)R_z(-\delta)e^{i(\psi'+\mu)/2}\ket{-}\\
    &= e^{i\psi'/2}\ket{u}.
\end{aligned}
\end{equation}
There is hence always a $G_R$ such that:
\begin{equation}
     R_x(\pi) G_R = U,
\end{equation}
for any $U \in \mathrm{SU}(2)$, and choosing $U = R_x(\pi)V$ for some $V \in \mathrm{SU}(2)$ shows that that $G_R$ is universal.
\section{Waveplate angle calculation}
In this section we give an explicit method for determining the waveplate angles $\alpha$, $\theta$ and $\phi$ in the reciprocal gadget given an arbitrary unitary $U\in \mathrm{SU(2)}$:
\begin{equation}
    G_R = Q(\theta) H(\phi) 
    H\Big(\frac{\pi}{8}\Big)
    F_-
    Q\Big(\frac{\pi}{2}\Big)
    H(\alpha)
    Q\Big(\frac{\pi}{2}\Big)
    F_+
    H\Big(\frac{\pi}{8}\Big) H(-\phi)Q(-\theta) = U.
\end{equation}
A Python implementation of the algorithm can be found in the online repository~\cite{MTQgit}. The method essentially consists of going through the proof presented in the previous section backwards. First, define a new unitary $V$:
\begin{equation}
    V = R_x(\pi) U,
\end{equation}
then find the eigenphase $-\lambda/2$ and corresponding eigenvector $\ket{v_{+}}$:
\begin{equation}
    V\ket{v_{+}} = e^{-i\lambda/2}\ket{v_{+}}.
\end{equation}
The angles $\gamma$ and $\delta$ should then be chosen to rotate $\ket{v_{+}}$ to $\ket{+}$. This can be done by taking:
\begin{align}
    \gamma &= \mathrm{arctan2} \Big(
    \mathrm{tr}\bigl[Z\ketbra{v_{+}}{v_{+}}\bigr]
    ,\mathrm{tr}\bigl[X\ketbra{v_{+}}{v_{+}}\bigr] \Big)\\
    \delta &= -\mathrm{arctan2} \Big(
    \mathrm{tr}\bigl[Y R_y(\gamma)\ketbra{v_{+}}{v_{+}}R_y^{\dagger}(\gamma)\bigr]
    ,\mathrm{tr}\bigl[X R_y(\gamma)\ketbra{v_{+}}{v_{+}}R_y^{\dagger}(\gamma)\bigr] \Big).
\end{align}
The angle $\psi$ is simply given by:
\begin{equation}
\psi = \lambda - \pi.
\end{equation} Next, the rotation angles $\gamma$ and $\delta$ need to be mapped to the corresponding waveplate angles $\theta$ and $\phi$:
\begin{equation}
    Q(\theta)H(\phi) = R_y(\gamma) R_z(\delta)R_x(\pi/2).
\end{equation}
To find these waveplate angles, first construct the state:
\begin{equation}
    \ket{L'} = R_x(-\pi/2)R_z(-\delta)R_y(-\gamma)\ket{L},
\end{equation}
where:
\begin{equation}
    R_x(-\pi/2)R_z(-\delta)R_y(-\gamma) = \bigl( Q(\theta)H(\phi) \bigr)^{-1}.
\end{equation}
The quarter-wave plate angle for the order $H(\phi')Q(\theta') = \pm R_y(\gamma) R_z(\delta)R_x(\pi/2)$ can then be found as:
\begin{equation}
    \theta' = \frac{1}{2} \mathrm{arctan2} \Big(
    \mathrm{tr}\bigl[X\ketbra{L'}{L'}\bigr]
    ,\mathrm{tr}\bigl[Z\ketbra{L'}{L'}\bigr]\Big) + \frac{\pi}{4}.
\end{equation}
To find the half-wave plate angle, construct the state:
\begin{equation}
    \ket{H'} = Q(\theta')R_x(-\pi/2)R_z(-\delta)R_y(-\gamma)\ket{H}.
\end{equation}
The angle $\phi'$ is then given as:
\begin{equation}
    \phi' = \frac{1}{4}\mathrm{arctan2} \Big(
    \mathrm{tr}\bigl[X\ketbra{H'}{H'}\bigr]
    ,\mathrm{tr}\bigl[Z\ketbra{H'}{H'}\bigr]\Big).
\end{equation}
Note that the ambiguity in the overall sign of the unitary doesn't matter due to palindromic order of the total gadget, since $\pm H(\phi')Q(\theta') = \pm ZQ(-\theta')H(-\phi')Z$, and any minus signs cancel. The angles for the order $Q(\theta)H(\phi)$ are found using the waveplate permutation rule:
\begin{equation}
    H(\alpha)Q(\beta) = Q(2\alpha-\beta)H(\alpha),
\end{equation}
and hence:
\begin{align}
    \phi &= \phi'\\
    \theta &= 2\phi - \theta'.
\end{align}
The last angle, the one of the middle half-wave plate, can be calculated directly from $\psi$:
\begin{equation}
    \alpha = \psi / 4 + \pi/2.
\end{equation}
Using the waveplate reduction rules~\cite{simon1990minimal}:
\begin{align}
    Q(a)H(b)H(c) &= Q(a+\pi/2)H(a-b+c-\pi/2)\\
    H(a)H(b)Q(c) &= H(a-b+c-\pi/2)Q(c+\pi/2),
\end{align}
the gadget can be simplified, removing two waveplates:
\begin{equation}
    G_R = Q(\theta_1) H(\phi_1) 
    F_-
    Q\Big(\frac{\pi}{2}\Big)
    H(\alpha)
    Q\Big(\frac{\pi}{2}\Big)
    F_+
    H(\phi_2)Q(\theta_2) = U,
\end{equation}
with:
\begin{align}
    \theta_1 &= \theta + \pi/2 \\
    \phi_1 &= \theta - \phi + \pi/8 - \pi/2\\
    \theta_2 &= -\theta + \pi/2\\
    \phi_2 &= \pi/8 + \phi - \theta - \pi/2.
\end{align}
%\clearpage
\section{Definitions of $G^{[,]}$, $G^{\{,\}}$}

The commuting and anti-commuting subsets $\mathcal{G}_+$, $\mathcal{G}_-$ of

\begin{equation}
    \mathcal{G} = \left\{ \mathbb{1}, X, Y, Z, 
    \frac{X+ Y}{\sqrt{2}}, \frac{X- Y}{\sqrt{2}}
    \frac{X+ Z}{\sqrt{2}}, \frac{X- Z}{\sqrt{2}}
    \frac{Y+ Z}{\sqrt{2}}, \frac{Y- Z}{\sqrt{2}} \right\}
\end{equation}
are
\begin{align}
    \begin{split}
        \mathcal{G}_+ &=\bigl\{ U_i, V_j \in \mathcal{G} | [U_i, V_j]=0 \bigr\} \\
        &= \bigl\{ U_i, V_j \in \mathcal{G} | i=0 \lor i=j \lor j=0 \bigr\}
    \end{split}
\end{align}
and
\begin{equation}
    \begin{alignedat}{5}
        \mathcal{G}_- &=\Bigl\{ U_i, V_j \in \mathcal{G} | &&\{U_i, V_j\}=0 \Bigr\}\\
         &=\Bigl\{ U_i, V_j \in \mathcal{G} |   &&i=1, j\in\{2,3,8,9\} \lor \\
         &\quad                 &&i=2, j\in\{1,3,6,7\} \lor \\
         &\quad                 &&i=3, j\in\{1,2,4,5\} \lor \\
         &\quad                 &&i=4, j\in\{3,5\}     \lor \\
         &\quad                 &&i=5, j\in\{3,4\}     \lor \\
         &\quad                 &&i=6, j\in\{2,7\}     \lor \\
         &\quad                 &&i=7, j\in\{2,6\}     \lor \\
         &\quad                 &&i=8, j\in\{1,9\}     \lor \\
         &\quad                 &&i=9, j\in\{1,8\} \Bigr\}.
    \end{alignedat}
\end{equation}
%\clearpage
\subsection{Supplementary data}
In this section we present alternative visualisations of the data presented in the main text, as well as some supplementary data used to generate the main result. 

\begin{figure*}[t]
        \includegraphics[width=1.00\linewidth]{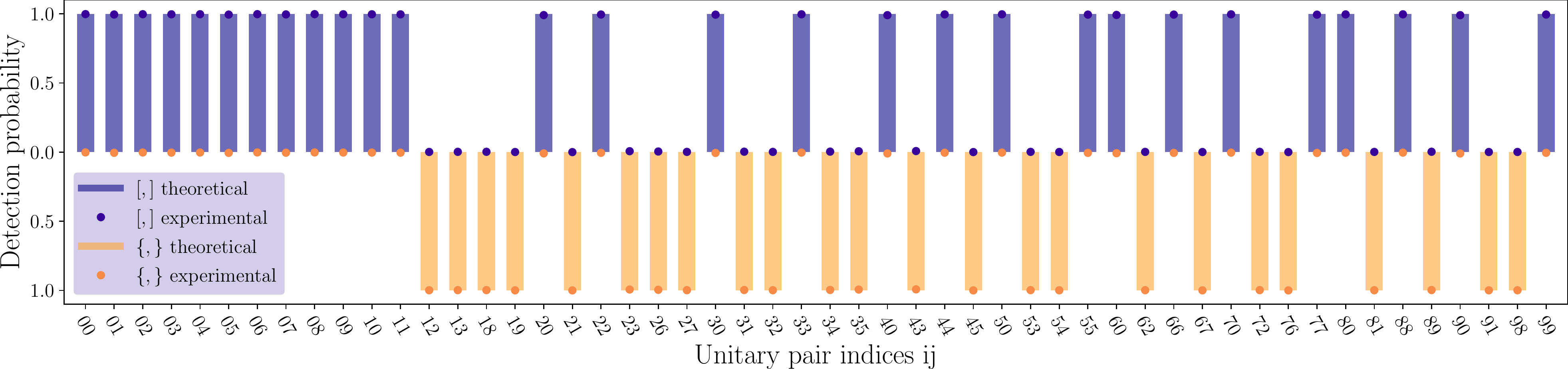}
        \centering
        \caption{\textbf{Average winning probabilities.} The figure shows the relative probability of a photon to be detected in the commutator and anti-commutator ports of the quantum SWITCH, for every pair of $(U,V)$ in the sets $G^{[,]}$ and $G^{\{,\}}$. The theoretical probabilities $p\in\{0,1\}$ are shown as solid bars, and the experimentally recorded ones, averaged over all six runs, are indicated by the colored dots. The indices $(i,j)$ on the x-axis specify the pair of unitary operators $(U,V)=(\mathcal{G}_i,\mathcal{G}_j)$.}
        \label{fig:histogram}
    \end{figure*}

\figref{fig:histogram} shows the expectation values for each pair of unitaries in the witness averaged over all runs. 

% In \figref{fig:histogram} the wining probabilities for each pair of unitaries in the witness is shown for every run of the experiment in a magnified scale. It can be seen that the results between rounds are very consistent, and the overall variance in the winning probability is small.

\begin{figure}
            \includegraphics[width=0.5\linewidth]{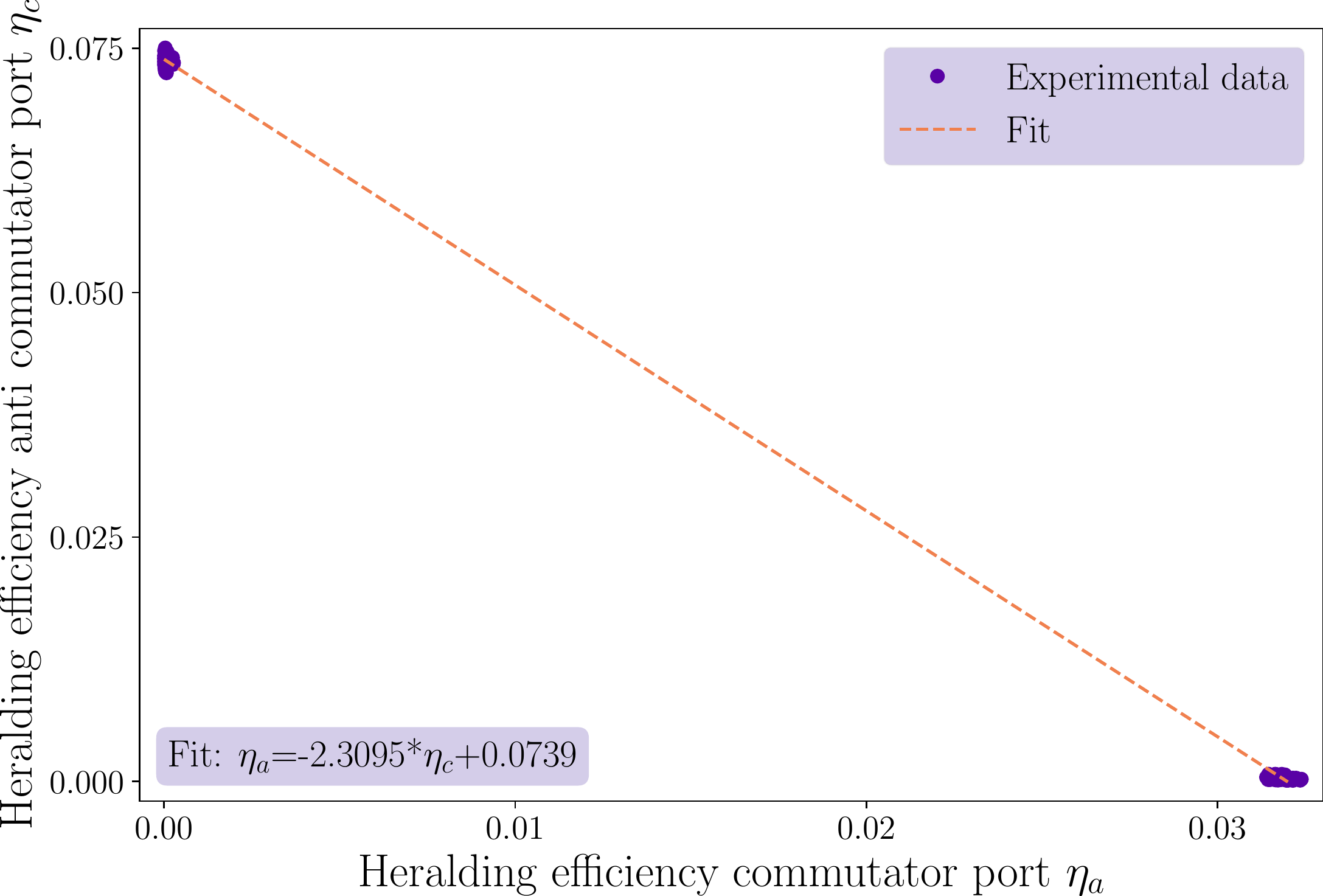}
            \centering
            \caption{\textbf{Relative heralding efficiencies}. Imbalanced detection efficiencies, as well as loss, between the two output modes of the quantum SWITCH may influence the calculated winning probabilities. As photons must exit the SWITCH in one of either ports, the observed relative heralding efficiencies of unitaries that commute to various degrees follow a linear slope.}
            \label{fig:detector_efficiency}
        \end{figure}

\begin{figure}
            \includegraphics[width=0.5\linewidth]{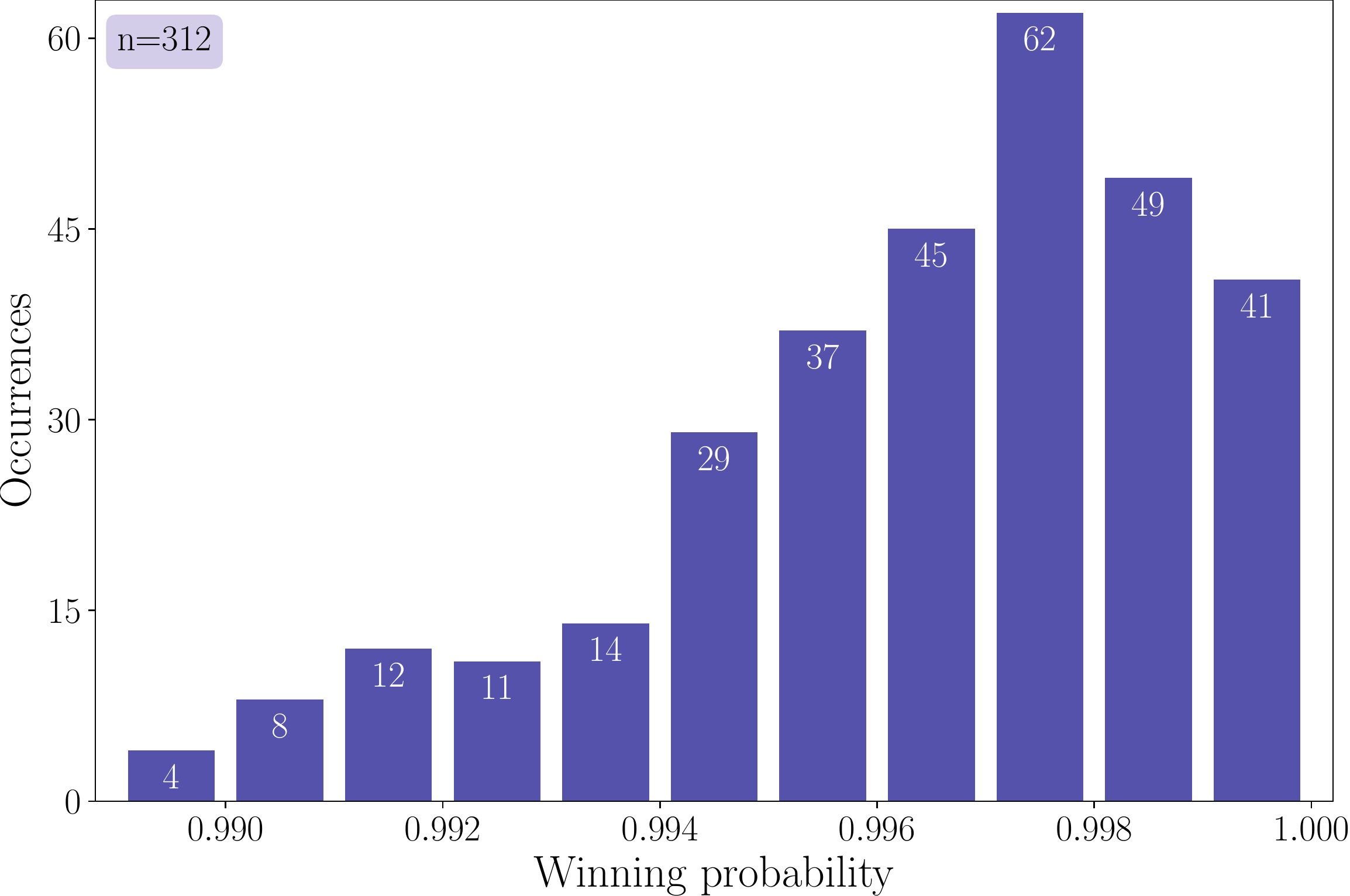}
            \centering
            \caption{Winning probability histogram (3 runs: n=312 )}
            \label{fig:histogram_winningprobability}
        \end{figure}

\figref{fig:detector_efficiency} shows the heralding efficiency in the $\ket{\pm}$ ports of the control qubit. Since the total photon number should be conserved, a linear fit to this data gives the relative detection efficiencies in the two ports. This value was in turn used in the evaluation of the success probabilities.. 
\figref{fig:histogram_winningprobability} shows a histogram of the winning probabilities with a bin size of $0.001$. This data includes all the settings for the six different runs. It can be seen that in the majority of rounds the winning probability exceeds $0.996$.

\begin{figure*}
            \includegraphics[width=0.8\linewidth]{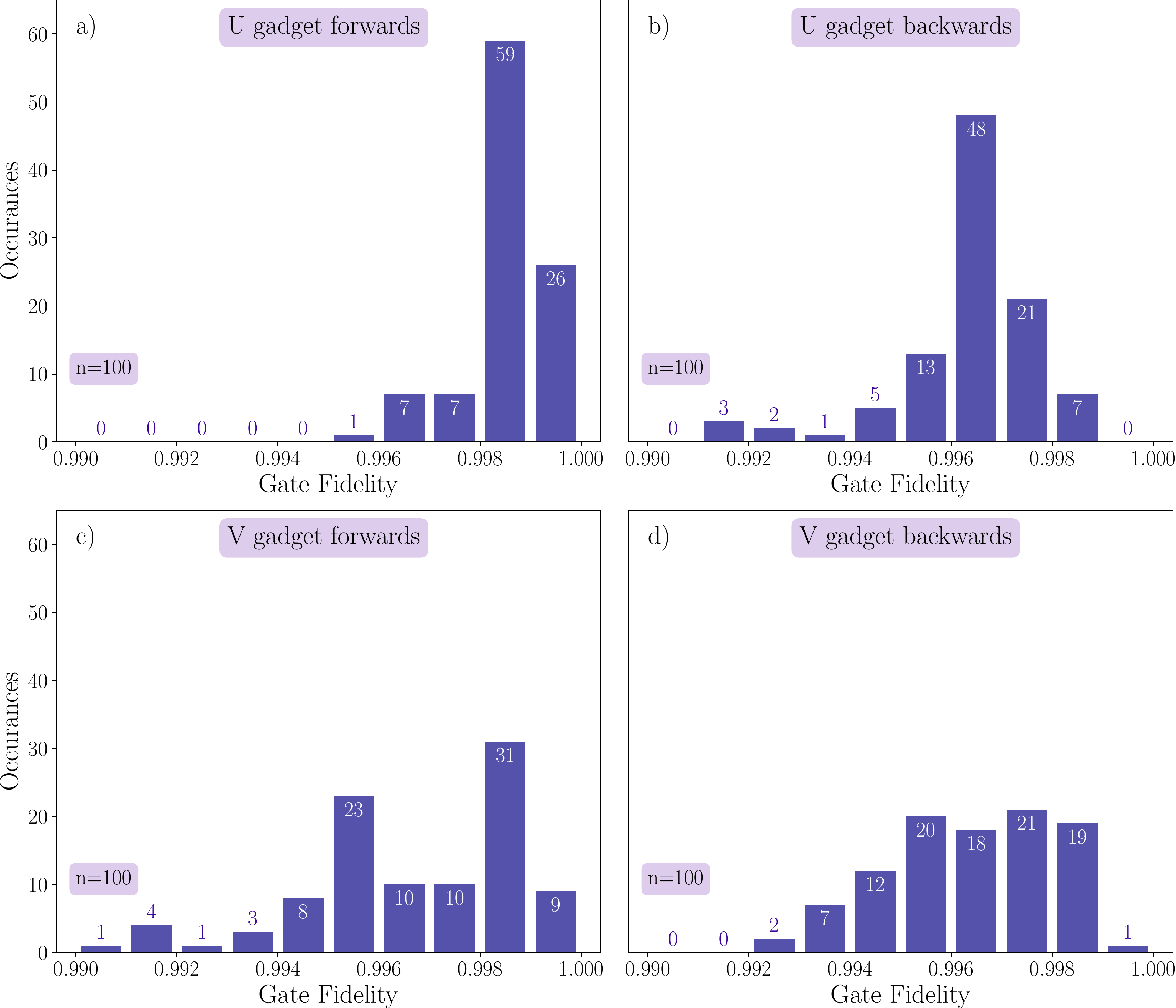}
            \centering
            \caption{Gate fidelity histograms for both gadgets in the forwards and backwards directions. 100 random unitaries $W_i$ were implemented with each gadget, the states $W_i\ket{\mathrm{H}}$ and $W_i\ket{\mathrm{+}}$ were measured in both directions independently and the quantum state fidelity between the expected and measured state determined. Depicted are the average fidelities for $\ket{H}$ and $\ket{+}$ for \textbf{a)} $U_{\mathrm{fw}}$ (Mean fidelity $0.99852\pm0.00079$). \textbf{b)} $U_{\mathrm{bw}}$ (\num{0.9964\pm0.0014}). \textbf{c)} $V_{\mathrm{fw}}$ (\num{0.9967\pm0.0021}). \textbf{d)} $V_{\mathrm{bw}}$ (\num{0.9964\pm0.0016}). The mean fidelity over all gadgets and directions is \num{0.9970\pm0.0018}. All uncertainties are standard deviations.}
            \label{fig:histogram_gate_fidelity}
        \end{figure*}

To verify that the gadgets can faithfully implement any unitary, we implement 100 random unitaries $W_i$, use quantum state tomography to determine the states $W_i^{\delta}\ket{\mathrm{\Psi}}$ for $\ket{\Psi}\in \left\{ \ket{\mathrm{H}}, \ket{\mathrm{+}} \right\}$ and $\delta \in \left\{ \mathrm{fw}, \mathrm{bw} \right\}$, and calculate the quantum state fidelities to the expected states. We define the gate fidelity as the average over $\Psi$, and show the resulting fidelities as histograms in \figref{fig:histogram_gate_fidelity}. The mean gate fidelity over all gadgets and directions of \num{0.9970\pm0.0018} indicates that the gadgets are indeed capable of implementing arbitrary unitaries.

\begin{figure}
            \includegraphics[width=0.8\linewidth]{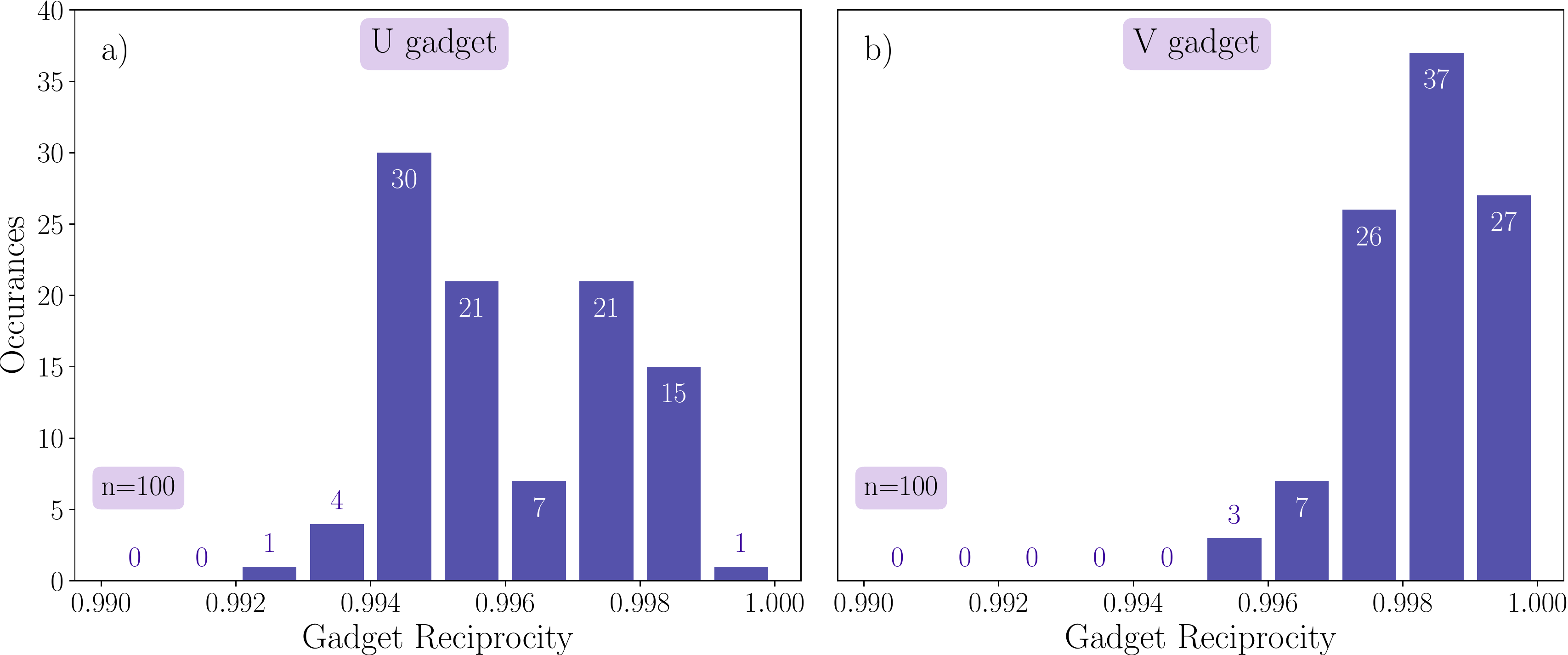}
            \centering
            \caption{Reciprocity histograms for the two gadgets used in the experiment. 100 random unitaries $W_i$ were implemented on both gadgets, the states $W_i\ket{\mathrm{H}}$ and $W_i\ket{\mathrm{+}}$ were measured in both directions independently and the quantum state fidelity between the expected and measured state determined. Depicted are the average fidelities for \textbf{a)} the U gadget, with mean fidelity of \num{0.9960\pm0.0016} and \textbf{b)} the V gadget, with a mean fidelity of $0.99834\pm0.00099$. The mean fidelity for both gadgets is \num{0.9972\pm0.0018}. All uncertainties are standard deviations.}
            \label{fig:histogram_gadget_reciprocity}
        \end{figure}

Finally, to check that the gadgets are reciprocal, we use the very same measurements, but now calculate the fidelities between $W_i^{\mathrm{fw}}\ket{\mathrm{\Psi}}$ and $W_i^{\mathrm{bw}}\ket{\mathrm{\Psi}}$. Similarly to before, we define the gadget reciprocity as an average over $\Psi$, and show the results in \figref{fig:histogram_gadget_reciprocity}. With a mean reciprocity of \num{0.9972\pm0.0018}, it can be concluded that unitaries implemented by the gadgets are reciprocal. We would like to point out that all unitaries were implemented independently for each direction, hence the reciprocity is affected by imperfect repeatability of the rotation motors moving the gadgets' waveplates.

\begin{figure}[t]
        \includegraphics[width=1.00\linewidth]{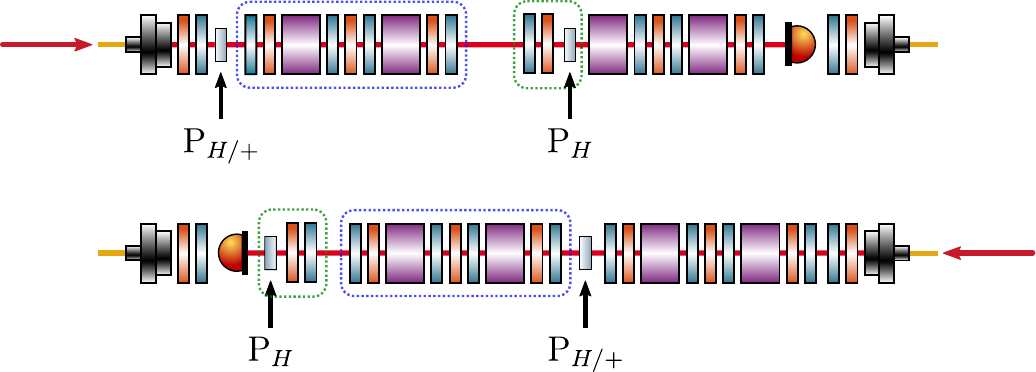}
        \centering
        \caption{\textbf{Tomography setup.} The tomography on the polarization unitaries was performed inside the Sagnac interferometer in order to avoid introducing measurement errors caused by imperfect polarization compensation in optical fibers. In the forward direction (top) the tomography on the first gadget was performed using the first two waveplates of the second gadget. A polarizer ($\mathrm{P_{H/+}}$) preparing the input polarization $H$ / $+$ was placed before the gadget, and a horizontally aligned measurement polarizer ($\mathrm{P_{H}}$) was placed inside the second gadget. In the backwards propagation direction (bottom) the state preparation polarizer was placed between the two gadgets, and two additional motorized waveplates were introduced into the setup to perform the tomography. The gadget being measured is circled in blue, while the measurement apparatus is circled in green. The tomography on the second gadget in forward (backward) direction was carried out analogously to the one on the first gadget in backward (forward) direction.}
        \label{fig:tomo}
    \end{figure}

In order to remove the influence of unwanted fiber polarization rotations on the tomography, it was not performed using the same polarization measurement stations as in the actual experiment. Instead, the tomography was carried out inside the Sagnac interferometer itself. In one propagation direction this was facilitated by using part of one polarization gadget to set the measurement basis for the tomography on the other gadget, and in the opposite propagation direction two additional motorized waveplates were introduced into the setup. A sketch of the tomography setup is shown in \figref{fig:tomo}.

%\clearpage

\subsection{Causal witness}
To characterise the causal structure of quantum processes one can make use of the process matrix formalism~\cite{oreshkov2012quantum}, in which a quantum process is represented by a positive semidefinite matrix $W$. The set of all process matrices that represent a definite causal structure, also called causally separable processes matrices, form a convex subset of all process matrices. Consequently, it is always possible to find a hyperplane separating any causally indefinite process matrix from the set of causally separable ones~\cite{Araujo2014}. This in turn implies the existence of a witness operator $\mathcal{S}$, that can be used to certify the indefinite causal structure of a process. Such a witness has previously been used experimentally to validate the casual non-separability of the quantum SWITCH~\cite{rubino2017experimental}. Here, we adapt a version of the witness from ~\cite{Araujo2014}, which is inspired by the task presented in Ref.~\cite{chiribella2012perfect}. Making use of the Choi-Jamiołkowski ismorphism, which allows us to represent linear maps and quantum channels as matrices, the witness can be defined in terms of the operators:
\begin{equation}
    G_{\pm}^{i,j} = \doubleket{U_i}\doublebra{U_i}\otimes
    \doubleket{V_j}\doublebra{V_j}\otimes\ket{\pm}\hspace{-2pt}\bra{\pm}_C,
\end{equation}
with:
\begin{equation}
    U_i,V_j \in \mathcal{G} = \left\{ \mathbb{1}, X, Y, Z, \frac{X\pm Y}{\sqrt{2}}, \frac{X\pm Z}{\sqrt{2}}, \frac{Y\pm Z}{\sqrt{2}} \right\}
    \label{eq:gset_apdx}.
\end{equation}
We then define
\begin{align}
     &\mathcal{G}_+:=\left\{\left(U_i,V_j\right) \Big| U_i,V_j \in \mathcal{G}, \; U_iU_j =V_jU_i  \right\} \\
     &\mathcal{G}_-:=\left\{\left(U_i,V_j\right) \Big| U_i,V_j \in \mathcal{G}, \; U_iV_j =-V_jU_i  \right\}
\end{align}
The witness itself is given by:
\begin{equation}
\label{eq:witnes}
    \mathcal{S} = \frac{1}{N}\sum_{i,j}^{10} q_{ij}^{[,]} G_+^{i,j} + 
    q_{ij}^{\{,\}} G_-^{i,j},
\end{equation}
where $q_{ij}^{[,]}$ and $q_{ij}^{\{,\}}$ are weights chosen such that:
\begin{equation}
    \begin{aligned}
    \{(i,j) : [U_i,V_j] = 0 \} &: g_{ij}^{[,]} = 1\\
    \{(i,j) : \{U_i,V_j\} = 0 \} &: g_{ij}^{\{,\}} = 1,
    \end{aligned}
\end{equation}
and are zero otherwise. Here, similarly to the main text, $N=52$ is the total number of commuting or anti-commuting pairs of unitaries in $\mathcal{G}$, and the coefficients above select exactly these subsets. The expectation value of the witness is evaluated as $\langle \mathcal{S} \rangle = \mathrm{tr}[\mathcal{S}W]$, where $W$ is a process matrix~\cite{oreshkov2012quantum}.
The causal separability bound for the witness described above can be evaluated numerically using the semidefinite programming methods of Ref.~\cite{Araujo2014}. Additionally, the methods from~\cite{bavaresco21} allow us to obtain a computer assisted proof that $\mathrm{tr}[\mathcal{S} W_{\mathrm{sep}}] \leq \frac{90.4}{100}$, for any causally separable process $W_{\mathrm{sep}}$; the code to certify this value is openly available in an online repository~\cite{MTQgit}. It can be shown that $\mathrm{tr}[\mathcal{S} W_{\mathrm{SWITCH}}] = 1$, where $W_{\mathrm{SWITCH}}$ is the process matrix of the quantum SWITCH.
    
\end{document}